# Vacuum nanogap formation in multilayer structures by an adhesion-controlled process

Z. Taliashvili[a], A. Tavkhelidze[b*], L. Jangidze[a], Y. Blagidze[c]

[a] *Tbilisi State University, Chavchavadze Ave. 13, 0179 Tbilisi, Georgia*

[b] *Ilia State University, Cholokashvili Ave. 3/5, 0162 Tbilisi, Georgia*

[c] *Institute of Cybernetics, S. Euli St. 5, 0186 Tbilisi, Georgia*

*E-mail address: avtotav@gmail.com

**Abstract**

In this study, we regulate adhesion between thin metal films to produce a large-area vacuum nanogap for electron tunneling. Multilayer structures comprising thin metal films with adjustable adhesion were fabricated. The Cu/Ag/Ti/Si structures were grown on Si substrates and include thin Ti and Ag films and a thick Cu layer. The Ag and Ti films were deposited on the Si substrate under vacuum, and a thick Cu layer was subsequently electroplated onto the Ag surface. Later, the sandwich was separated and a vacuum nanogap was opened to produce two Ag/Cu and Ti/Si conformal electrodes. The adhesion strength between the Ti and Ag films was precisely adjusted by exposing the structures to dry $O_2$ after Ti growth but before Ag growth. The resulting adhesion needed to be sufficiently high to allow electroplating of Cu and sufficiently low to allow subsequent separation. Either heating or cooling was used to separate the sandwiches. T he structures separated as a result of the different thermal expansion parameters of the Si and Cu electrodes and the low adhesion between the Ti and Ag layers. After separation, the Ag and Ti surfaces were analyzed optically using a Michelson interferometer. Adhesion regulation and optimization of the electroplating regime allowed fabrication of two conformal electrodes with a nanogap smaller than 5 nm and an area larger than 7 $mm^2$. Such electrodes can be used for efficient energy conversion and cooling in the mixed thermionic and thermotunneling regime.

## 1. Introduction

In recent years, micro-coolers and energy converters based on electron tunneling have been investigated [1–6]. The earlier study of cooling by electron tunneling was conducted to prevent overheating in single-electron transistors [1]. A typical metal-insulator-metal (MIM) tunnel junction has high heat conductivity as a result of its low insulator thickness, which is in the range of 3–10 nm. This creates parasitic heat backflow, which decreases the cooling coefficient of an MIM junction. A metal-vacuum-metal (MVM) junction is free from this drawback but is notably complex and difficult to realize practically. However, a conformal electrode method based on the electroplating of a top electrode has been proposed [2]. Coolers utilizing vacuum nanogaps have been studied in a number of earlier publications [3–7]. It has been shown that the cooling power of MVM junctions is high, on the order of 100 W/$cm^2$ [3, 5]. However, the cooling coefficient of such junctions does not exceed 10%. A metal-vacuum-insulator-metal (MVIM) tunnel junction with an additional thin layer of insulator coating has also been studied [6]. It was found that the cooling coefficient of this MVIM junction was higher (40–50%) than that of an MVM junction, and, additionally, the insulating layer protects the electrodes against the possibility of a short circuit and simplifies the design. Composite electrode materials were also studied from the point of view of maximizing thermionic emission and thermotunneling [8]. The practical realization of energy converters and coolers based on thermotunneling has been discussed in the literature [9–16]. Multilayer structures allowing the formation of vacuum nanogaps on a large scale has been reported in reference [17].

The objective of this work is to develop a method that allows precise regulation of the adhesion between Ti and Ag films in Cu/Ag/Ti/Si multilayer structures. This adhesion must be sufficiently high to allow the electroplating of Cu and sufficiently low to allow the subsequent separation (i.e., separation of the electrodes without the need for a solid barrier) without significant deformation of the Ag/Cu and Ti/Si bulk electrodes. The matching between the two conformal electrodes should be accurate enough to allow electron tunneling in the MVM junction formed between the electrodes.

**2. Experimental details**

*2.1 Sandwich growth*

Cu/Ag/Ti/Si structures were grown on a Si substrate and include thin films of Ti and Ag and a thick Cu layer. The Ag and Ti films were deposited on a Si substrate under vacuum and the thick Cu layer was subsequently electroplated. We used n-type Si (100) double-sided polished substrates with diameters of 50 mm, 40 mm, and 20 mm, thicknesses in the range of 1–2 mm and surface roughnesses of less than 0.5 nm as the base electrodes. The Cu/Ag/Ti/Si sandwiches were fabricated according to the following steps:

(i) A Si substrate was exposed to a chemical treatment procedure consisting of wet cleaning in dimethylformamide at $T=325$ K for 3 minutes; wet cleaning in a mixture of HF and $H_2O$ (1:25, volume fraction) for 30 seconds; plasma cleaning in a vacuum chamber using an $O_2$ plasma (at an $O_2$ pressure of 13 Pa) for 20 minutes, followed by an Ar plasma (at an Ar pressure of 13 Pa) for 20 minutes; and, finally, heating to $T_S=470$ K prior to deposition. The Si substrate temperature was measured with a precision of 0.5 K.

(ii) A thin film of Ti with a thickness in the range of 80–100 nm was deposited via thermal evaporation of Ti (99.99%) at $P=2\times10^{-4}$ Pa over a period of 2–3 seconds in a spiral-shaped tungsten evaporator (substrate temperature $T_S=440$ K).

(iii) The film was exposed to dry $O_2$ for a period of 0–40 seconds at $P=(4–10)\times10^{-4}$ Pa.

(iv) A thin Ag film with a thickness of approximately 1.2 μm was deposited via thermal evaporation of Ag (99.99%) at $P=2.5\times10^{-4}$ Pa over a period of 2–3 seconds (substrate temperature $T_S =440$ K).

(v) A layer of Cu 0.5–4 mm in thickness was electroplated using a sulfate electrolyte ($CuSO_4 \times 5H_2O + H_2SO_4 + C_2H_5OH + H_2O$) in a thermo-stabilized bath at a current density of $J=15–50$ mA/cm$^2$ under mechanical stirring.

(vi) Chemical polishing of Cu was carried out (in a mixture of $H_3PO_4$, $HNO_3$, and $CH_3COOH$ at a ratio of 25:50:25, volume fractions) at $T=308$ K for 1.5 minutes.

Magnetron sputtering was also used to grow the Ti and Ag films. In this case, steps (ii)–(iv) above were replaced by the following steps:

(ii*) DC magnetron sputtering of Ti (99.99%) was performed at a magnetron voltage of 370 V with a current of 1.5 A for 15 seconds at $P=7.9\times10^{-1}$ Pa and a substrate temperature of $T_S=423–433$ K.

(iii*) The resulting film was exposed to dry $O_2$ for a period of 0–40 seconds at a pressure in the range of $8\times10^{-4}$ to $8\times10^{-3}$ Pa.

(iv*) DC magnetron sputtering of Ag (99.99%) was carried out at a voltage of 450 V with a current of 0.5 A over 80 seconds at $P=8\times10^{-1}$ Pa and a substrate temperature of $T_S=425–450$ K.

The sample was subsequently exposed to ambient air and a thick Cu layer was electroplated on the Ag surface. The Cu layer was deposited using a sulfate electrolyte ($CuSO_4 \times 5H_2O + H_2SO_4 + C_2H_5OH + H_2O$) in a thermo-stabilized bath, at a current density in the range of $J=15–50$ mA/cm$^2$ under mechanical stirring. Two baths were used to stabilize the electrolyte temperature. The electrolyte temperature was varied within the range of $T=296.5–308$ K and stabilized with an accuracy of 0.3 K. The electrolyte used was simple in its composition, stable, easy to manipulate and allowed high current densities. Ethyl alcohol was used to prevent the formation of monovalent Cu ions, resulting in dense, finely crystalline precipitates. An electrically insulated current lead was connected to the Si wafer from the backside.

*2.2 Sandwich separation*

Cu/A g/Ti/Si sandwiches (Fig. 1a) were subsequently separated by using the reduced adhesion between the Ti and Ag layers. Bulk Ti/Si and Ag/Cu electrodes are shown in Fig. 1b. Either heating or cooling was used to separate the sandwiches. The structures separated as a result of the different thermal expansion parameters of the Si and Cu electrodes and the low adhesion between the Ti and Ag layers. The separations took place in the process of heating within the temperature range of 298–360 K and cooling within the range 298–260 K.

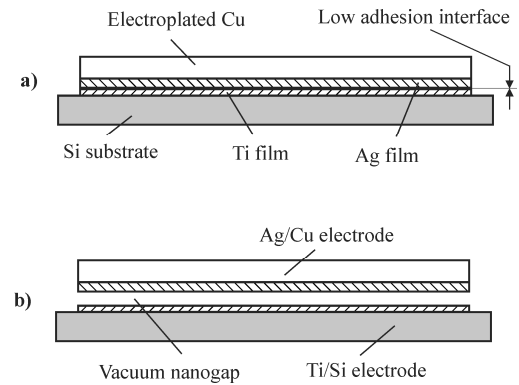

Fig. 1 Cross-section of a Cu/Ag/Ti/Si sandwich: a) before separation: b) and after separation .

Two numbers are used in this work to describe adhesion: the probability of separation damage, $q_{sd}$, which describes probability of damage in the process of separation (gauge damage to the Si substrate); and the probability of electroplating damage, $q_{ed}$, which describes the probability of liquid entering the Ag/Ti interface during electroplating. These two probabilities were used instead of the tape-pull test to estimate adhesion. After separation, the thicknesses of the Ag/Cu electrodes were measured using a point micrometer (Heidenhain).

The best samples were separated by heating or cooling inside an evacuated volume. For this process, each sandwich was placed inside a PZT-type cylindrical piezoelectric actuator produced by PI Ceramic, which had electrodes on the inside and outside surfaces for applying voltage. This cylindrical actuator was used for vacuum-gap opening and adjustment, and also as housing. After vacuum was reached inside the housing, the sandwich was separated. DC voltage was then applied to the actuator electrodes to open the vacuum nanogap in situ. Nanogap width was adjusted by applying varying DC voltages in the range of -700 V to +500 V to the piezoelectric actuator electrodes.

*2.3 Surface conformity measurements*

The surface conformity (i.e., the absolute curvature value) of the Ag/Cu electrodes was measured using a Michelson interferometer oriented toward the Ag side [18]. An He-Ne laser with a wavelength of $\lambda$=632.8 nm served as a light source. One of the interferometer mirrors was replaced by the mirror-surface of an Ag/Cu electrode. An interference pattern was formed in the air gap between one of the mirrors and a virtual image of the second mirror (in our case, the sample). When the air gap was plane-parallel, fringes of equal inclination (circular fringes) were obtained (Fig. 2). After fixing, the sample was adjusted to obtain an interference pattern with fringes of equal inclination. By counting the number of fringes, $n$, the absolute curvature could be calculated using the formula $N = n \times \lambda / 2$. The sign of the curvature was examined by pressing the longer leg of the mirror. An unbending of the fringes in the interference pattern indicated surface convexity and a bending of the fringes indicated concavity. To determine a normalized curvature $\alpha$ for the metal electrode, the formula $\alpha = n (\lambda / 2) / D$ was used. The initial normalized curvatures for the base electrodes (Si wafers) of diameter 50 mm, 40 mm, and 20 mm were 57 nm/mm, 47 nm/mm, and 16 nm/mm, respectively. We monitored the temperature during the interferometer measurements to ensure that it was equal to the electroplating temperature.

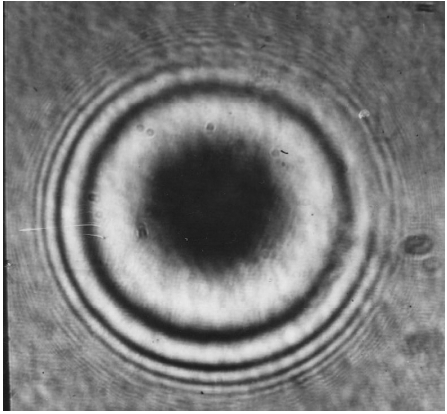

Fig. 2 Michelson interference rings for an Ag/Cu electrode (observed from the Ag side). The light source was a He–Ne laser of wavelength 630 nm.

**3. Results and Discussion**

The initial samples fabricated without exposure to $O_2$ exhibited mechanically damaged Ti/Si and Ag/Cu surfaces after separation. The Ti/Si electrode exhibited gouging damage on the Si wafer and the Ag/Cu electrode was strongly deformed. This damage resulted from the high adhesion between the thin films. It was difficult to obtain sandwiches with predefined separation temperatures. The results in general were not reproducible, but some dependence upon the residual gas pressure $P_r$ was noted. Additionally, it was found that results were dependent on the time interval between the Ti and Ag depositions, $\tau$, and substrate temperature, $T_S$. When the parameters $P_r$, $T_S$ and $\tau$ were varied, structures with variable adhesions were obtained. Therefore, we introduce formally three intervals of adhesion: (i) strong adhesion, if gouging damage to the Si wafer was observed after separation; (ii) intermediate adhesion, if the structures separated without visible damage; (iii) weak adhesion, if electrolyte entered the Ti/Ag interface during the electroplating. To describe the adhesion numerically we calculate several parameters, defined as follows. The probability of separation damage was calculated as $q_{sd} = (N_{sd} / N_{total})$, where $N_{sd}$ is the number of separation-damaged (gouging-damaged) samples and $N_{total}$ is the total number of samples. The probability of electroplating damage was calculated as $q_{ed} = (N_{ed} / N_{total})$, where $N_{ed}$ is the number of samples in which electrolyte entered the Ag/Ti interface and $N_{total}$ is the total number of samples. The total number of samples in all calculations was $N_{total} = 300$.

*3.1 Adhesion control*

To regulate adhesion more precisely, we introduced the step of flooding the deposition chamber with dry $O_2$. Samples were exposed to $O_2$ after Ti deposition until a given pressure $P_{O2}$ was obtained, and the $O_2$ was then pumped out before Ag deposition. Fig. 3 presents variations in $q_{sd}$ and probability of electroplating without damage $1 - q_{ed}$ (or 1 minus the probability of electroplating damage $q_{ed}$) as functions of $P_{O2}$ for substrate temperatures in the range of $T_S$=440 K and an $O_2$ exposure time of $\tau$=30 seconds. To present the pressure dependence of adhesion, we combine the probability functions $q_{sd}$ and $1 - q_{ed}$ and plot them together.

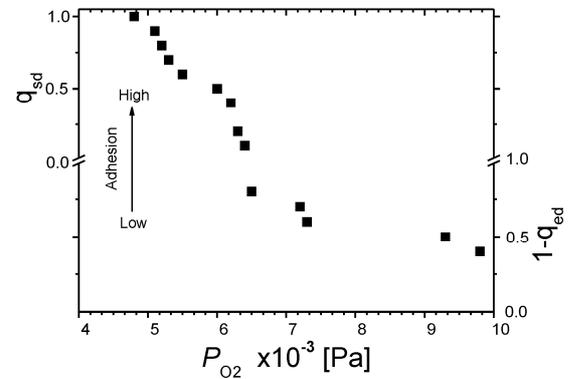

Fig. 3 Probabilities of separation damage, $q_{sd}(P)$, and electroplating without damage, $1 - q_{ed}(P)$, where $q_{ed}(P)$ is the probability of electroplating damage, expressed as a function of $O_2$ pressure between growth of the Ti and Ag layers. Data recorded at constant $T_S$ =440 K and $\tau$=20 seconds.

Axis breaks were introduced to separate the two functions. As shown in Fig. 3, adhesion decreases with increasing $P_{O2}$. The highest-quality Ti/Si and Ag/Cu $q_{sd} \cong 0$, $q_{ed} \cong 0$, corresponding to a pressures in the range of $P_{O2} = 5.5 \times 10^{-3}$ – $6.5 \times 10^{-3}$ Pa. Within this range, both damage probabilities, $q_{sd}$ and $q_{ed}$, were low.

We also investigated the dependence of the probabilities $q_{sd}$ and $1 - q_{ed}$ on the exposure time, $\tau$, with the substrate temperature and O$_2$ pressure held constant at $T_S = 440$ K and $P_{O2} = 6 \times 10^{-3}$ Pa. These results are shown in Fig. 4. The data show that adhesion decreases with increasing $\tau$.

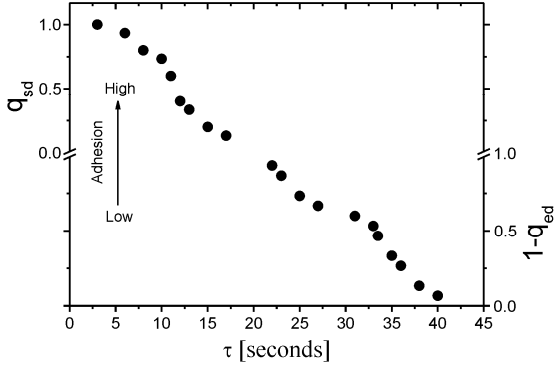

Fig. 4 Probabilities of separation damage, $q_{sd}(P)$, and electroplating without damage, $1 - q_{ed}(P)$, where $q_{ed}(P)$ is the probability of electroplating damage, expressed as a function of O$_2$ exposure time between growth of the Ti and Ag layers. Data recorded at constant $T_S = 440$ K and $P_{O2} = 6 \times 10^{-3}$ Pa.

Finally, the dependence of the probabilities $q_{sd}$ and $1 - q_{ed}$ on the substrate temperature was recorded, with the O$_2$ pressure and exposure time held constant at $P_{O2} = 6 \times 10^{-3}$ Pa and $\tau = 20$ second (Fig. 5). The data show that adhesion increases with increasing substrate temperature.

No difference in adhesion regulation was observed between the thermally evaporated and sputtered films of Ti and Ag. The sandwiches incorporating both sputtered films and thermally evaporated ones separated at approximately the same temperatures and the separated electrodes showed quite similar relative curvatures.

The above experimentally determined adhesion results can be interpreted according to the following scheme. Exposing a Ti film to O$_2$ during fabrication oxidizes the surface of the Ti film, forming an ultra-thin TiO$_2$ layer with a reduced surface energy. The self-cleaning properties of TiO$_2$ layers are described in reference [19]. This reduction in surface energy is characteristic of all metal oxides, including Ti [20]. The surface energy is reduced further as the oxide coverage increases. After the entire surface is covered with an oxide layer, the surface energy continues to decrease with increasing oxide thickness. This assumption follows from consideration of the driving force of the continued reaction as the thickness of the oxide layer increases. After the added O$_2$ is pumped out, an Ag film is deposited, covering the TiO$_2$ layer. Because Ag does not react with TiO$_2$ [21], there are no chemical bonds at the TiO$_2$/Ag interface. The ultra-thin TiO$_2$ layer thus serves as a barrier between the Ti and Ag atoms, preventing their intermixing. The TiO$_2$ layer simultaneously makes the Ti surface chemically inert. It follows that the Ti/Ag interface has the lowest adhesion with respect to the other interfaces (Ti/Si, Ag/Cu) in the sandwich. Consequently, the sandwich separates at the TiO$_2$/Ag interface. The interactions between TiO$_2$ films and Ag have been studied previously [22, 23]. According to this model, if adhesion is reduced with increasing oxide coverage, it should also be reduced with increasing oxide thickness after complete surface coverage is achieved. This is supported by the data from Figs. 3 and 4. An increase in oxide thickness and a corresponding decrease in adhesion are observed in both figures. In the case of Fig. 3, the increase in oxide thickness results from increasing O$_2$ pressure at fixed $T_S$ and $\tau$, while in the case of Fig. 4, this increase in oxide thickness results from increasing $\tau$ at fixed $P_{O2}$ and $T_S$. In a high-adhesion regime (Figs. 3, 4) with a high probability of separation damage, the TiO$_2$ coverage of the surface appears to be incomplete. Ti and Ag atoms intermix within open areas on the surface (i.e., areas where there is no TiO$_2$ barrier), forming a Ti-Ag alloy. This mixing leads to high local adhesion and separation damage. In the low-adhesion regime, TiO$_2$ coverage is complete and the oxide film is thick, resulting in an adhesion that is sufficiently low for electrochemical liquid to enter between the TiO$_2$ and Ag films. The intermediate adhesion regime between the two regimes mentioned above was most suitable for sandwich separation without electrode damage and deformation.

The observed increase in adhesion with increasing substrate temperature at fixed $P_{O2}$ and $\tau$ (Fig. 5) can be explained by the presence of adsorbed O$_2$ on the TiO$_2$ surface. Adsorbed O$_2$ is in thermodynamic equilibrium with the gas phase and its surface concentration is reduced with increasing surface temperature. Adhesion therefore increases with increasing $T_S$. This result is also supported by the higher rate of adhesion increase with temperature for higher temperatures, i.e., $T_S$>430 K (Fig. 5).

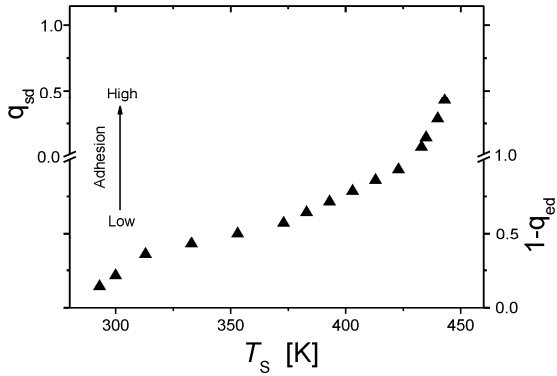

Fig. 5 Probabilities of separation damage, $q_{sd}(P)$, and electroplating without damage, $1 - q_{ed}(P)$, where $q_{ed}(P)$ is the probability of electroplating damage, expressed as a function of temperature during exposure to O$_2$ between growth of the Ti and Ag layers. Data recorded at constant $P_{O2} = 6 \times 10^{-3}$ Pa and $\tau = 20$ seconds.

To obtain conformal electrodes, we chose the fabrication conditions $P_{O2}$ =6x10$^{-3}$ Pa, $T_S$ =400 K and $\tau$=20 seconds as the optimal technological parameters. This optimization resulted in a stable separation temperature and a reduced normalized curvature for the Ag/Cu electrode.

*3.2 Electrode conformity*

A fundamental problem under investigation was the surface conformity of the electrodes. Initially, a majority of the Si substrates were convex (as a result of the mechanical polishing step). Each Si wafer was observed to curve in an arbitrary direction with respect to its initial curvature during the process of Cu electroplating . After separation , each wafer always returned to its initial curvature. However, the Ag/Cu electrode adopted a concave (Fig. 6a) or convex (Fig. 6b) curvature depending on the electroplating regime . A concave Ag/Cu electrode was needed to obtain a conformal pair. We investigated the dependence of the sign of the curvature (i.e., the fraction of concave and convex surfaces expressed as a percentage) on the wafer thickness and electroplating temperature . When wafers of 1-mm thickness were used, the percentage of concave surfaces was 22%. For 2-mm thick wafers, the percentage of concave surfaces was 59%. Increasing the wafer thickness thus increased the percentage of concave surfaces by a factor of 2.7. The dependence of the sign of the curvature on the electrolyte temperature was investigated as well. The results for 2-mm thick wafers with Ag/Cu diameters of 12 mm and 3 mm are given in Table 1. The best results were obtained for $T$= 306 K and 308 K, respectively. More detailed data on electroplating is presented in reference [17].

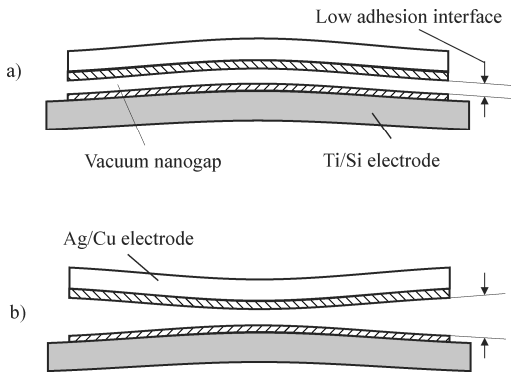

Fig. 6 Sandwich after separation in the cases of a) convex Ti/Si electrode and concave Ag/Cu electrode; b) convex Ti/Si electrode and convex Ag/Cu electrode.

Table 1. Percentage of Ag/Cu electrodes that were concave for several values of electroplating temperature. The electrode diameters were 12 mm and 3 mm.

| | Ag/Cu with $D$=12 mm | | | | | Ag/Cu with $D$=3 mm | |
|---|---|---|---|---|---|---|---|
| $T$ [K] | 296.5 | 299 | 303 | 306 | 308 | 296.5 | 308 |
| Concave surfaces [%] | 54 | 24 | 26 | 80 | 82 | 100 | 85.7 |

Compressive prestress was present in the Ag/Cu electrodes, exhibited by the fact that these electrodes were always more convex than the corresponding Si/Ti electrodes (Fig. 6). In the absolute majority of cases when electroplating failed, edge delamination of the Ag/Cu electrode was observed. Buckle delamination was observed only in a few test samples. These results need further explanation because in other experiments the opposite results have been observed (i.e., buckle delamination dominated over edge delamination) [24]. Our hypothesis is that a much lower value of compressive prestress $\sigma$ was present in our experiments in comparison to those previously reported (because our work was focused on the reduction of $\sigma$ [17]). This explanation is in good agreement with an edge delamination model for the compressed films. As shown in reference [25], the apparent edge toughness that is observed in most experiments vanishes for low values of $\sigma$.

*3.3 Tunneling area evaluation*

We also evaluated the tunneling area, defined as the interior region for which the distance between the electrodes, $a$, was less than 5 nm. The tunneling area can be estimated from simple geometric calculations. For these calculations, we consider the cross-section of the sandwich (Fig. 7). The tunneling area is defined as the area of the ring with an external radius equal to the sample radius OC and an internal radius OF. Replacing the arcs AC and BC with straight lines, we find that the contact area can be expressed by the formula $S = \pi r^2 (a/H)(2 - a/H)$ with adequate accuracy. Within this expression, $H$=AB is the maximum difference between the curvatures and $a$=DE. Table 2 presents results for the samples with 3-mm diameters. $H$ was determined from the expression $H = n \times \lambda$, where $n$ is fringe number.

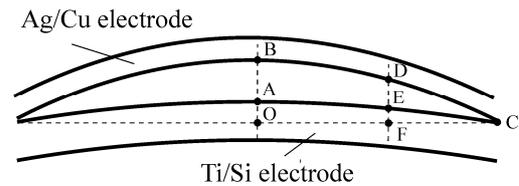

Fig. 7 Representation of the sandwich cross-section, where AE is the Ti/Si electrode effective surface and BD is the Ag/Cu electrode effective surface.

Table 2. Michelson interference rings and corresponding contact areas for 3-mm-diameter Ag/Cu electrodes grown at $T$=296.5 K and $T$=308 K.

| $T$=296.5 K | | $T$=308 K | |
|---|---|---|---|
| $n$ [ring count] | $S$ [mm$^2$] | $n$ [ring count] | $S$ [mm$^2$] |
| <0.025 | 6–7 | <0.025 | 6–7 |
| <0.025 | 6–7 | <0.025 | 6–7 |
| 0.17 | 1.22 | 0.17 | 1.22 |
| 0.04 | 4.42 | 0.04 | 4.42 |
| <0.025 | 6–7 | <0.025 | 6–7 |
| 0.025 | 6.05 | 0.025 | 6.05 |

Table 2 shows that the maximum tunneling areas were obtained for samples grown at $T$=296.5 K. In this case, the area of the entire interior region was tunneling area, signifying that a vacuum nanogap of width less than 5 nm was formed (for an electrode 3 mm in diameter).

**4. Conclusions**

Multilayer Cu/Ag/Ti/Si sandwiches were grown to obtain vacuum nanogap tunnel junctions. The adhesion at the Ag/Ti interface was controlled to separate each sandwich and open the vacuum nanogap between the Ti/Si and Ag/Cu bulk electrodes. Adhesion was reduced by pumping dry $O_2$ into the vacuum chamber between the growth of the Ti and Ag thin films. A thick Cu electrode was then electroplated on the Ag film and the structure was separated by heating or cooling using a thermal expansion mismatch. Adhesion at the Ag/Ti interface was optimized by varying the $O_2$ pressure $P_{O2}$, the substrate temperature $T_S$ and the $O_2$ exposure time $\tau$. The probabilities of separation damage and electroplating damage were recorded for 300 samples as functions of $P_{O2}$, $T_S$ and $\tau$. Based on these experimental results we propose that an ultra-thin $TiO_2$ film grows on Ti during its exposure to $O_2$. This $TiO_2$ layer reduces the surface energy and also prevents Ti and Ag atoms from intermixing at the Ag/Ti interface. The adhesion reduction depends on the $TiO_2$ coverage in the case of an ultra-thin $TiO_2$ layer and on the $TiO_2$ film thickness in the case of complete coverage. The optimal technological parameters were found to be $P_{O2}$=6x10$^{-3}$ Pa, $T_S$=400 K and $\tau$=20 seconds. The Ag/Cu and Ti/Si electrode surfaces were analyzed via Michelson interferometry and both the adhesion strength and electroplating regime were adjusted to obtain conformal electrodes for vacuum nanogap tunnel junction applications.

**Acknowledgments**

The authors thank A. Bibilashvili, G. Skhiladze and R. Melkadze for useful discussions. This work was partly supported by Borealis Technical.

**References:**

[1] A.N. korotkov, M.R. Samuelsen, S.A. Vasenko, J. Appl. Phys. 76 (1994) 3623- 3631.
[2] A. Tavkhelidze, G. Skhiladze, A. Bibilashvili, L. Tsakadze, L. Jangidze, Z. Taliashvili, I. Cox, Z. Berishvili, Proc. XXI International Conf. on Thermoelectricsc, August 26–29 IEEE, New York, pp. (2002) 435–438.
[3] Y. Hishinuma, T. H. Geballe, B. Y. Moyzhes, and T. W. Kenny, *Appl*. Phys. Lett. 78 (2001) 2572.
[4] T. Zeng, Appl. Phys. Lett., 88 (2006) 153104.
[5] Z. Xin, Z. Dian-Lin, Chinese Phys. 16 (2007) 2656–2660.
[6] A. Tavkhellidze, V. Svanidze, L. Tsakadze, J. Vac. Sci. Technol. A, 26 (2008) 5.
[7] Hishinuma I., Geballe T.H., Moyzhes B.I., Kenny T.W., Appl. Phys. Lett. 94 (2003) 4690.
[8] A. N. Tavkhelidze, J. Appl. Phys. 108 (2010) 044313.
[9] E.A. Chávez-Urbiola, Yu.V. Vorobiev, L.P. Bulat, Solar Energy 86 (2012) 369–378.
[10] G. B. Smith and C. G. Granqvist, Green Nanotechnology, Solutions for Sustainability and Energy in the Built Environment, CRC Press, (2011).
[11] C. Knight and J. Davidson, in Advances in wireless sensors and sensor networks, S.C. Mukhopadhyay, H. Leng (Ed.), Springer, (2010), p. 221
[12] L. P. Bulat, D. A. Pshenaĭ-Severin, Physics of the Solid State 52 (2010) 485–492.
[13] Y. C. Gerstenmaier, G. Wachutka, AIP Conf. Proc. 890 (2007) 349.
[14] G. Despesse, T. Jager, J. Appl. Phys. 96 (2004) 5026.
[15] H. J. Goldsmid, Thermoelectrics, La Grande Motte, France, August 17 21, 2003, Proceedings ICT'03. 22nd International Conference on Thermoelectrics (2003) 433.
[16] M. H. Tanielian, R. B. Greegor, J. A. Nielsen, and C. G. Parazzoli, Appl. Phys. Lett., 99, 123104 (2011).
[17] L. Jangidze, A.Tavkhelidze, Y. Blagidze, Z. Taliashvili, J. Electrochem. Soc. 159 (2012) D413-D417.
[18] P. Hariharan, Optical interferometry, Elsevier, (2003).
[19] T. Tolkea, A. Kriltzb, A. Rechtenbacha, Thin Solid Films 518 (2010) 4242–4246.
[20] J. A. Venables, Introduction to surface and thin film processes, Cambridge university press, 2000, p. 4
[21] U. Diebold, Surface Science Reports 48 (2003) 53–223.
[22] S.-Y. Jung, T.-J. Ha, C.-S. Park, W.-S. Seo, Y. S. Lim, S. Shin, H. H. Cho, H.-H. Park, Thin Solid Films 529 (2013) 94-97.
[23] N. Duraisamy, N. M. Muhammada, H.-C. Kimb, J.-D. Joc, K.-H. Choi, Thin Solid Films 520 (2012) 5070–5074.
[24] L. B. Freund, S. Suresh, Thin Film Materials: Stress, Defect Formation and Surface Evolution, Cambridge University Press, 2003
[25] D. S. Balint, J. W. Hutchinson, J. Appl. Mech., 68 (2001) 725